\documentclass[preprint,tightenlines,showpacs,floatfix,nofootinbib,
superscriptaddress,fleqn]{revtex4}
\usepackage{graphicx}
\usepackage{graphicx}
\usepackage{bm}
\usepackage{dcolumn}
\usepackage{amsmath}
\usepackage{amsfonts}
\usepackage{amssymb}

\begin{document}

\bibliographystyle{prsty}


\preprint{PNU-NTG-11/2005}
\preprint{RUB-TPII-09/2005}

\title{Parity-violating aysmmetries in elastic $\vec{e}p$
  scattering in the chiral quark-soliton model: Comparison with A4, G0, HAPPEX and SAMPLE.}

\author{Antonio Silva}
\email{ajose@teor.fis.uc.pt}
\affiliation{Departamento de F\'\i sica and Centro de
F\'\i sica Computacional, Universidade de Coimbra,
P-3000 Coimbra, Portugal}
\affiliation{Faculdade de Engenharia da
Universidade do Porto, R. Dr. Roberto Frias s/n, P-4200-465 Porto,
Portugal}

\author{Hyun-Chul Kim}
\email{hchkim@pusan.ac.kr} \affiliation{Department of Physics and
Nuclear Physics \& Radiation Technology Institute (NuRI), Pusan
National University, 609-735 Busan, Republic of Korea}

\author{Diana Urbano}
\email{urbano@fe.up.pt} 
\affiliation{Faculdade de Engenharia da
Universidade do Porto, R. Dr. Roberto Frias s/n, P-4200-465 Porto,
Portugal}
\affiliation{Departamento de F\'\i sica and Centro de
F\'\i sica Computacional, Universidade de Coimbra,
P-3000 Coimbra, Portugal}

\author{Klaus Goeke}
\email{klaus.goeke@tp2.rub.de}
\affiliation{Institut f\"ur Theoretische  Physik II,
Ruhr-Universit\" at Bochum,  D--44780 Bochum, Germany}
\date{August 2005}

\begin{abstract}
We investigate parity-violating electroweak asymmetries in the
elastic scattering of polarized electrons off protons within the
framework of the chiral quark-soliton model ($\chi$QSM).  We use
as input the former results of the electromagnetic and strange
form factors and newly calculated SU(3) axial-vector form factors,
all evaluated with the same set of four parameters adjusted
several years ago to general mesonic and baryonic properties.
Based on this scheme, which yields positive electric and magnetic
strange form factors with a $\mu_s=(0.08-0.13)\mu_N$, we determine
the parity-violating asymmetries of elastic polarized
electron-proton scattering. The results are in a good agreement
with the data of the A4, HAPPEX, and SAMPLE experiments and
reproduce the full $Q^2$-range of the G0-data. We also predict the
parity-violating asymmetries for the backward G0 experiment.
\end{abstract}

\maketitle

{\bf 1.} The complex structure of the nucleon goes well beyond its
simplest description as a collection of three valence quarks
moving in some potential. The sea of gluons and $\rm
q\bar{q}$-pairs that arises in quantum chromodynamics is expected
to play an important role even at long distance scales. As the
lightest explicitely non-valence quark the strange quark provides
an attractive tool to probe the $\rm q\bar{q}$-sea, since any
strange quark contribution to an observable must be the effect of
the sea. Thus the strange quark contribution to the distributions
of charge and magnetization in the nucleon has been a very
important issue well over decades, since it provides a vital clue
in understanding the structure of the nucleon.  For recent
reviews, see, for example,
Refs.~\cite{Musolf:1993tb,Kumar:2000eq,Beck:2001dz,
Beck:2001yx,Ramsey-Musolf:2005rz}. Recently, the strangeness
content of the nucleon has been studied particularly intensively
 since parity-violating electron scattering (PVES) has demonstrated to
provide an essential tool for probing the sea of $\rm s\bar{s}$
pairs in the vector channel~\cite{Cahn:1977uu,Kaplan:1988ku}.  In
fact, various PVES experiments have been already conducted in
order to measure the parity-violating asymmetries (PVAs) from which
the strange vector form factors can be
extracted~\cite{Mueller:1997mt,Spayde:2000qg,SAMPLE00s,
Aniol:2000at,Maas,happex,Aniol:2005zf,Aniol:2005zg,Armstrong:2005hs}.
While PVES experiments have direct access to the PVA with
relatively good precision, a certain amount of uncertanties arise
in the flavor decomposition for the nucleon vector form factors.
As a result, the strange vector form factors extracted so far from
the data have rather large
errors~\cite{Mueller:1997mt,Spayde:2000qg,SAMPLE00s,
Aniol:2000at,Maas,happex,Aniol:2005zf,Aniol:2005zg}.

The chiral quark-soliton model ($\chi$QSM) is an effective quark
theory of the instanton-degrees of freedom of the QCD vacuum. It
results in an effective chiral action for valence and sea quarks
both moving in a static self-consistent Goldstone background
field~\cite{Christov:1996vm,Alkofer:1994ph} originating from the
spontaneous chiral symmetry breaking of the QCD. It has
successfully been applied to mass splittings of hyperons, to
electromagnetic and axial-vector form
factors~\cite{Christov:1996vm} of the baryon octet and decuplet
and to forward and generalized parton
distributions~\cite{Diakonov:1996sr,Petrov:1998kf, Goeke:2001tz}
and has led even to the prediction of the heavily discussed
pentaquark baryon $\Theta^+$~\cite{Diakonov:1997mm}. The present
authors have recently investigated in the $\chi$QSM model the
strange vector form factors~\cite{Silva:2001st,Silva:2002ej} and
they presented some aspects of the SAMPLE, HAPPEX, and A4
experiments. The results have shown a good agreement with the
available data, though the experimental uncertainties are rather
large, as mentioned above. Thus, it is theoretically more
challenging to calculate directly the PVAs and to confront them with
the more accurate experimental data.  Moreover, since the G0
experiment has measured the PVA over a range of momentum transfers
$0.12 \le Q^2 \le 1.0\,{\rm GeV}^2$ in the forward
direction~\cite{Armstrong:2005hs}, the check of the theory is on much
firmer ground. 

Actually, the PVA contains a set of six electromagnetic form
factors ($G_{E,M}^{u,d,s}$) and three axial-vector ones
($G_{A}^{u,d,s}$). In fact, all these form factors have already
been calculated within the
SU(3)-$\chi$QSM~\cite{Silva:2001st,Silva:2002ej,Silva,Silvaetal}
by using the well established parameter set consisting of $m_{s} =
180\,{\rm MeV}$ and the other three parameters having been adjusted some
years ago to the physical values of $f_{\pi} $, $m_{\pi} $ and
baryonic properties as e.g. the charge radius of the proton and
the delta-nucleon ($\Delta-N$) mass splitting. Apart from reproducing the
existing experimental data on the PVAs, we will predict the PVAs of
the future G0 experiment at backward angles.\\

\begin{figure}[ht]
  \centering
\includegraphics[height=7cm]{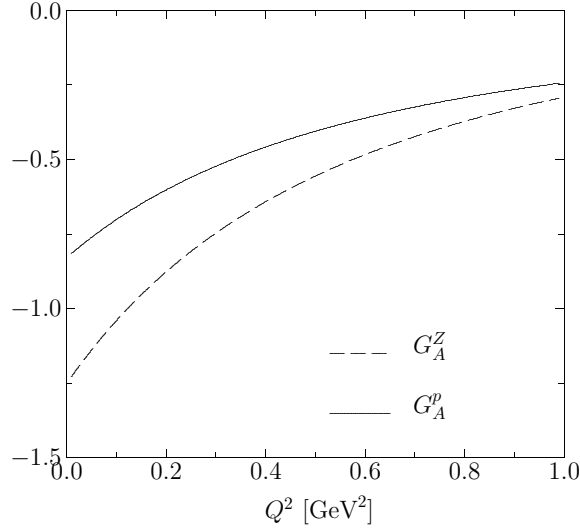}
\caption{The electroweak neutral axial-vector form factors $G_A^e$
and $G_A^{pZ}$ as functions of $Q^2$ calculated in the $\chi$QSM.}
  \label{fig:0}
\end{figure}

\vspace{0.5cm}

{\bf 2.} The PVA in polarized $\vec{e}p$ scattering is defined as
the difference of the total cross sections for circularly
polarized electrons with positive and negative helicities divided
by their sum:
\begin{equation}
\mathcal{A}_{PV}=\frac{\sigma_{+}-\sigma_{-}}{\sigma_{+}+\sigma_{-}}.
\label{eq:asym}
\end{equation}
Denoting, at the tree level, the amplitudes for $\gamma$ and $Z$ exchange
by $\mathcal{M}_\gamma$ and $\mathcal{M}_Z$, respectively, the total cross
section for a given polarization is proportional to the square of the
sum of the amplitudes, which indicates the interference between the
electromagnetic and neutral weak amplitudes:
\begin{equation}
\sigma_{\pm}\sim |\mathcal{M}^\gamma+\mathcal{M}^Z|^2_{\pm} .
\label{eq;cross}
\end{equation}
The PVA comprises three different terms:
\begin{equation}
\mathcal{A}_{PV} = \mathcal{A}_V+\mathcal{A}_s+\mathcal{A}_A,
\label{eq:asymmetry}
\end{equation}
where
\begin{eqnarray}
\mathcal{A}_V &=& -a \rho' \left[(1-4\kappa'
  \sin^2\theta_W) - \frac{\varepsilon  G_E^p G_E^n + \tau G_M^p
  G_M^n}{\varepsilon(G_E^p)^2 + \tau(G_M^p)^2}\right],\cr
\mathcal{A}_s &=& a\rho'\left[\frac{\varepsilon G_E^p
  G_E^s + \tau G_M^p G_M^s}{\varepsilon(G_E^p)^2 +
  \tau(G_M^p)^2}\right],\cr
\mathcal{A}_A &=& a  \left[\frac{(1-4\sin^2\theta_W)\varepsilon' G_M^p
  G_A^p}{\varepsilon(G_E^p)^2 +
  \tau(G_M^p)^2}\right],\cr
a &=& G_{F}Q^{2}/\left( 4\sqrt{2}\pi\alpha_{\rm
  EM}\right),\cr
\tau&=& Q^{2}/(4M_{N}^{2}),\cr
\varepsilon &=& \left[1+2(1+\tau)\tan^2\theta/2\right]^{-1},\cr
\varepsilon' &=& \sqrt{\tau(1+\tau)(1-\varepsilon^2)}.
  \label{eq:var}
\end{eqnarray}
The $G_{E,M}^p$, $G_{E,M}^s$, and $G_A^p$ denote, respectively,
the electromagnetic form factors of the proton, strange vector
form factors, and the axial-vector form factors.  The $G_F$ is the
Fermi constant as measured from muon decay, $\alpha_{\rm EM}$ the
fine structure constant, and $\theta_W$ the electroweak mixing
angle given as $\sin^2\theta_W=0.2312$~\cite{Eidelman:2004wy}. The
$Q^2$ stands for the negative square of the four momentum
transfer.  The parameters $\rho'$ and $\kappa'$ are related to
electroweak radiative corrections~\cite{Musolf:1993tb,Zhu:2000gn}.

\begin{figure}[h]
  \centering
\includegraphics[height=7cm]{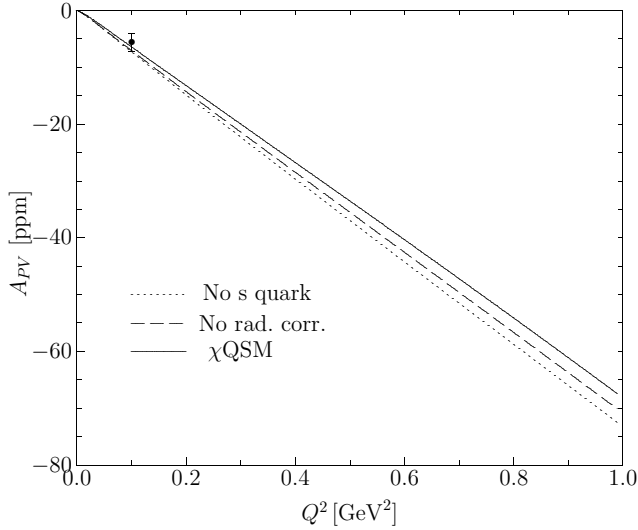}
  \caption{The parity-violating asymmetries as a function of $Q^2$,
compared with the SAMPLE measurement~\cite{Spayde:2000qg}. The
dotted curve is calculated without the s-quark contribution. The
dashed curve is obtained by using the form factors from the
$\chi$QSM without the electroweak radiative corrections, while the
solid one ($\chi$QSM) includes them and is our final result.}
  \label{fig:1}
\end{figure}

\begin{figure}[h]
  \centering
\includegraphics[height=7cm]{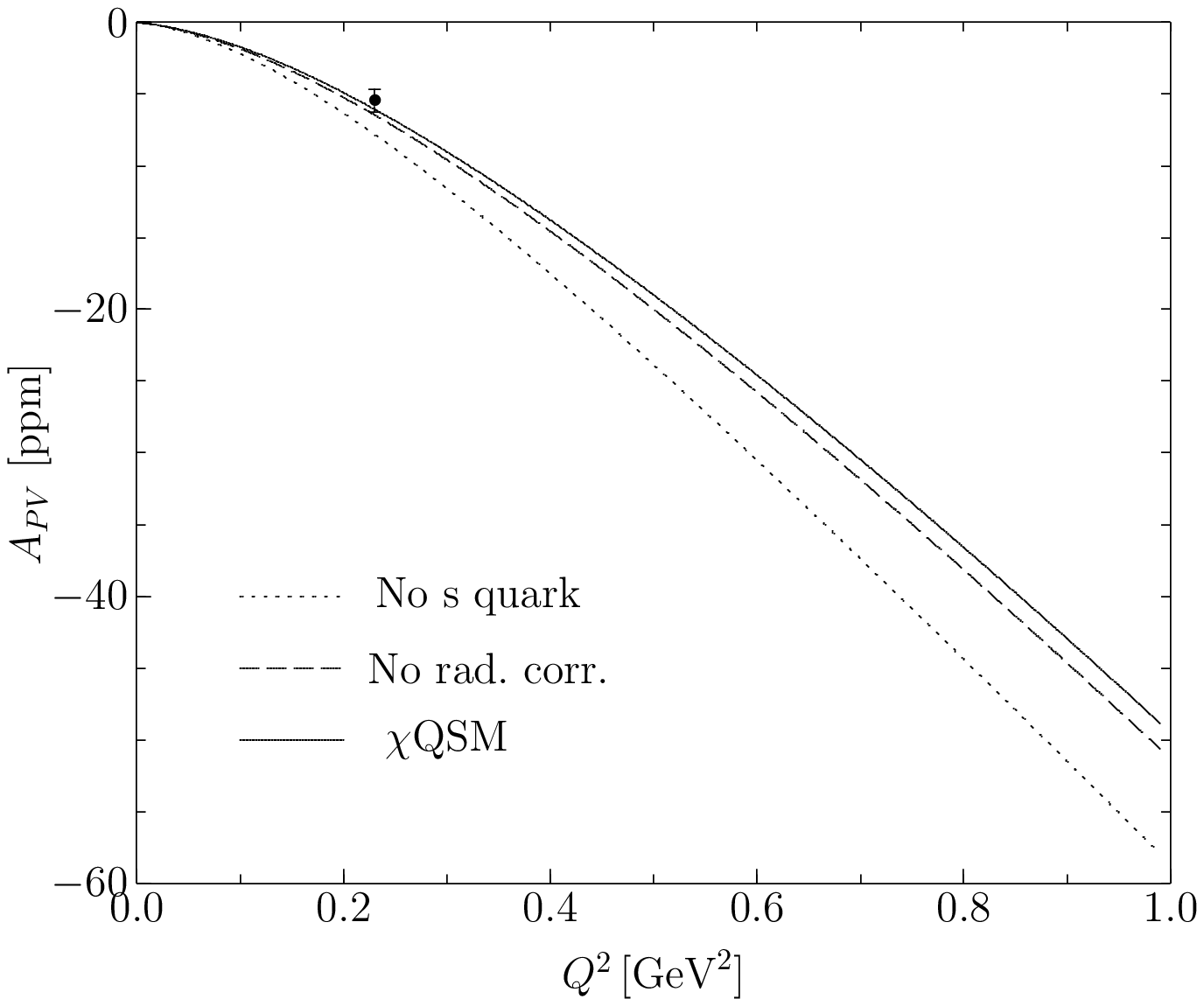}
  \caption{The parity-violating asymmetries as a function of $Q^2$,
compared with the A4 measurement~\cite{Maas}. The dotted curve is
calculated without the s-quark contribution. The dashed curve is
obtained by using the form factors from the $\chi$QSM without the
electroweak radiative corrections, while the solid one ($\chi$QSM)
includes them and is our final result.}
  \label{fig:2}
\end{figure}

\begin{figure}[h]
  \centering
\includegraphics[height=7cm]{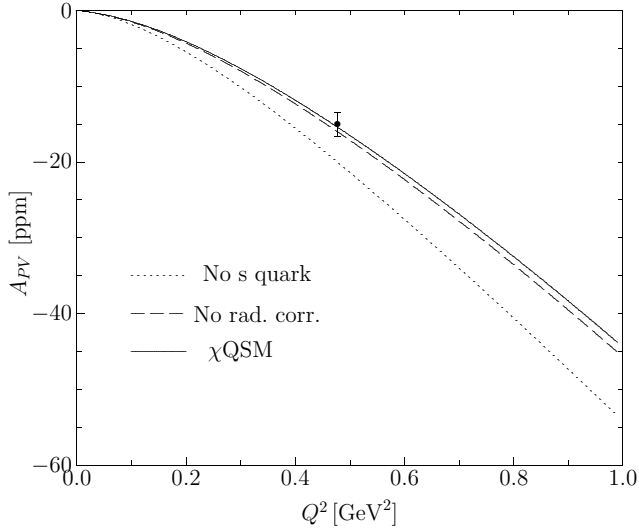}
  \caption{The parity-violating asymmetries as a function of $Q^2$,
compared with the HAPPEX measurement~\cite{Aniol:2000at}. The
dotted curve is calculated without the s-quark contribution. The
dashed curve is obtained by using the form factors from the
$\chi$QSM without the electroweak radiative corrections, while the
solid one ($\chi$QSM) includes them and is our final result.}
  \label{fig:3}
\end{figure}

Factoring out the quark charges, we can express the electromagnetic
and electroweak neutral axial-vector form factors of the proton in
terms of the flavor-decomposed electromagnetic form factors:
\begin{eqnarray}
G_{E,M}^p &=& \frac23 G_{E,M}^u - \frac13 \left(G_{E,M}^d +
G_{E,M}^s\right),\cr G_{A}^{pZ} &=& G_A^d -
\left(G_A^u+G_A^s\right). \label{eq:decomp}
\end{eqnarray}
Including the electroweak radiative
corrections~\cite{Musolf:1993tb,Zhu:2000gn},
we find that the electroweak axial-vector form factors of the proton
can be written as~\cite{Alberico:2001sd}:
\begin{equation}
G_A^p(Q^2)= - (1+R_A^1) G_A^{(3)}(Q^2) + R_A^0 + G_A^s,
\label{eq:GAe2}
\end{equation}
with the values for the electroweak radiative
corrections~\cite{Zhu:2000gn}:
\begin{equation}
 R_A^1=-0.41\pm0.24 ,\;\; R_A^0=0.06\pm0.14.
\end{equation}
Figure~\ref{fig:0} depicts the electroweak neutral axial-vector
form factors expressed in Eqs.(\ref{eq:decomp},\ref{eq:GAe2}),
which is obtained in the $\chi$QSM~\cite{Silvaetal}. We will use
$G_A^{pZ}$ in Fig.~\ref{fig:0} to yield the PVA.

The other six electromagnetic form factors, $G_{E,M}^{p,n,s}$ can
be read out from Refs.~\cite{Silva:2001st,Silva:2002ej,Silva}.

\vspace{0.5cm} {\bf 3.}  We discuss now the results of the PVA
obtained from the $\chi$QSM.  In detail, the model has the
following parameters: The constituent quark mass $M$, the current
quark mass $m_{\rm u}$, the cut-off $\Lambda$ of the proper-time
regularization, and the strange quark mass $m_{\rm s}$. However,
these parameters are not free but has been fixed to independent
observables in a very clear way~\cite{Christov:1996vm}: For a
given $M$ the $\Lambda$ and the $m_{\rm u}$ are adjusted in the
mesonic sector to the physical pion mass $m_\pi = 139$ MeV and the
pion decay constant $f_\pi = 93$ MeV. The strange quark mass is
selected to be $m_{\rm s} = 180$ MeV throughout the present work,
with which the mass splittings of hyperons are produced very well.
The remaining parameter $M$ is varied from $400$ MeV to $450$ MeV.
However, the value of $420\,{\rm MeV}$, which for many years is
known to produce  the best fit to many baryonic
observables~\cite{Christov:1996vm}, is chosen for our final result
in the baryonic sector.  We always assume isospin symmetry.  With
these parameters at hand, we can proceed to derive the form
factors of the proton required for the PVA.  On obtaining these
form factors, we use the symmetry conserving quantization scheme
\cite{Praszalowicz:1998jm} and take into account the rotational
$1/N_c$ corrections, the explicit SU(3) symmetry breaking in
linear order, and the wave function corrections, as discussed in
Ref.~\cite{Christov:1996vm,Silva:2001st} in detail. With this
scheme, we have obtained the
results~\cite{Silva:2001st,Silva:2002ej} for the strange vector
form factors in good agreement with the data of the A4, SAMPLE and
HAPPEX experiments as far as they were available\footnote{The
value of the strange electric form factor at $Q^2=0.091\, {\rm
GeV}^2$ is newly extracted by the HAPPEX
experiment~\cite{Aniol:2005zf}: $G_E^s =(-0.038\pm 0.042\pm
0.010)n.m.$ which is consistent with zero. The G0 experiment
indicates that $G_E^s$ may be negative in the intermediate region
up to $Q^2\sim 0.3\,{\rm GeV}^2$.  The present model predicts
$G_E^s\simeq 0.025$ at $Q^2=0.091\,{\rm GeV}^2$ which is positive
and slightly outside the error margins of HAPPEX.}.

\begin{figure}[h]
  \centering
\includegraphics[height=7cm]{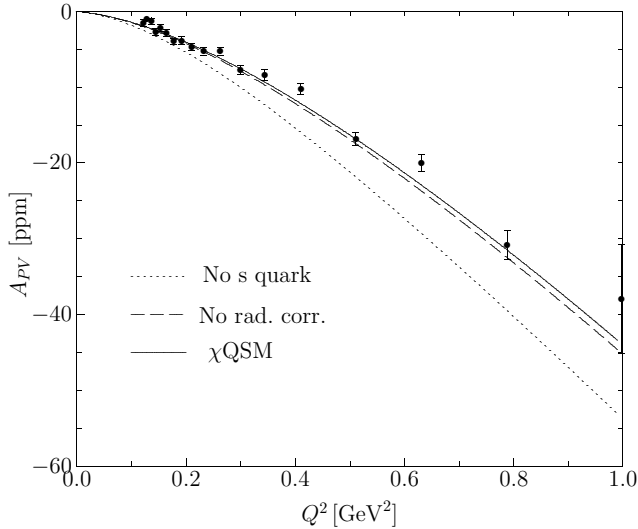}
  \caption{The parity-violating asymmetries as a function of $Q^2$,
compared with the forward G0 measurement~\cite{Armstrong:2005hs}.
The dotted curve is calculated without the s-quark contribution.
The dashed curve is obtained by using the form factors from the
$\chi$QSM without the electroweak radiative corrections, while the
solid one ($\chi$QSM) includes them and is our final result.
 }
  \label{fig:4}
\end{figure}

\begin{figure}[h]
  \centering
\includegraphics[height=7cm]{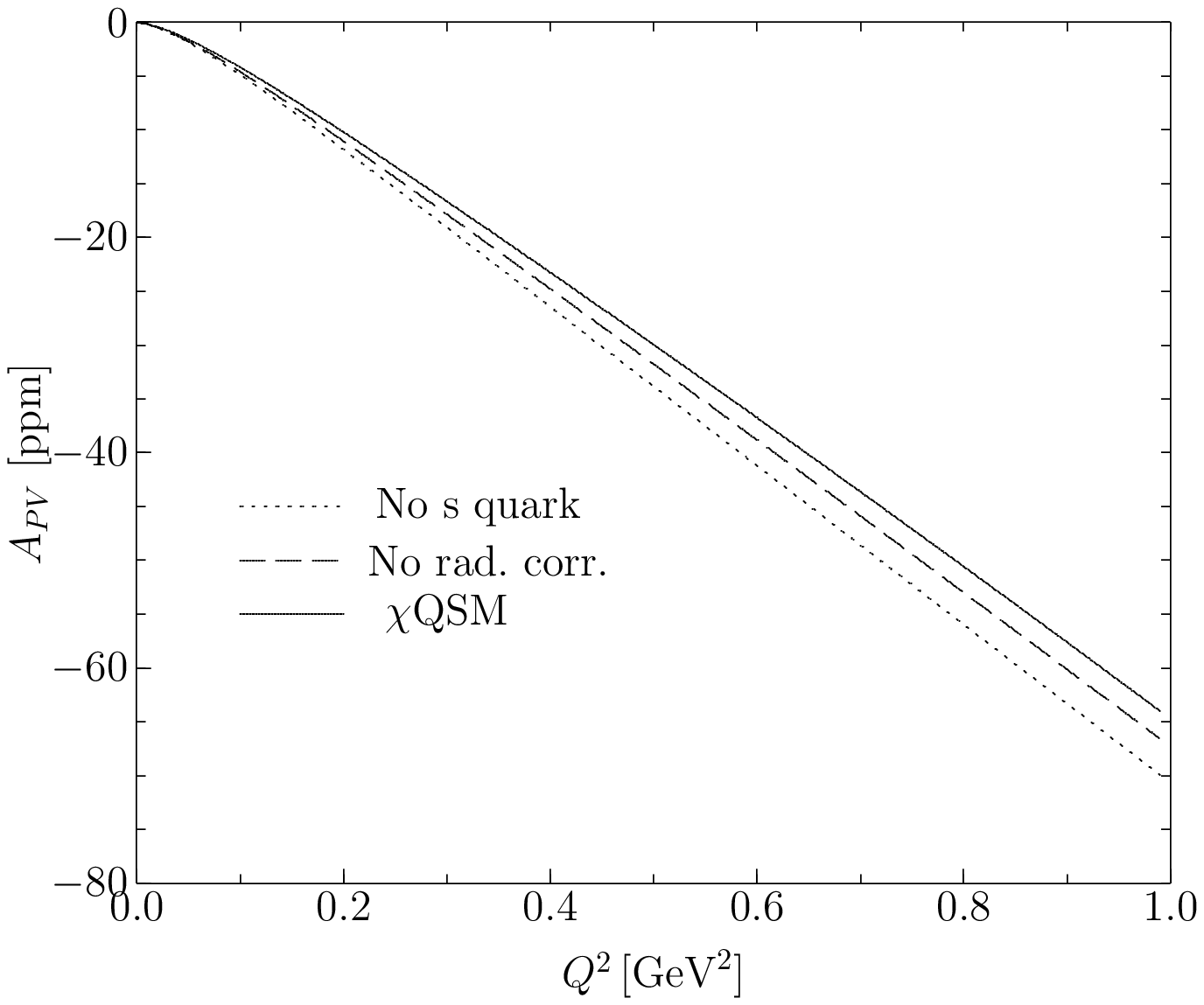}
  \caption{The parity-violating asymmetries as a function of $Q^2$.
  They are the predictions for the backward G0 experiment ($\theta =108$). The
dotted curve is calculated without the s-quark contribution. The
dashed curve is obtained by using the form factors from the
$\chi$QSM without the electroweak radiative corrections, while the
solid one ($\chi$QSM) includes them and is our final result.}
  \label{fig:5}
\end{figure}

\begin{figure}[h]
  \centering
\includegraphics[height=7cm]{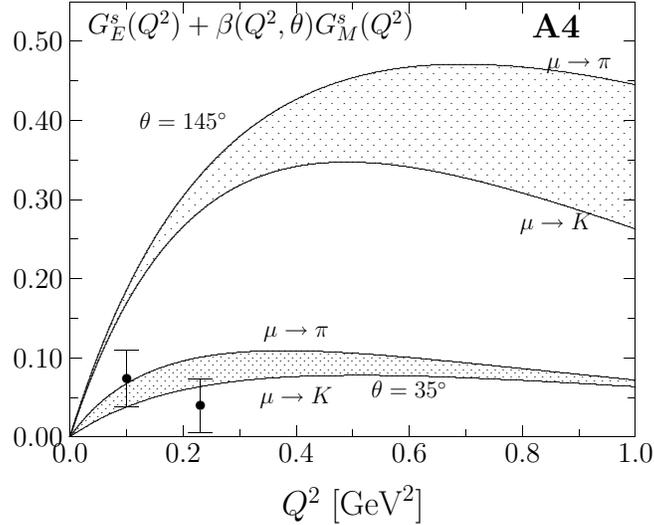}
  \caption{The values of $G_E^{s}(Q^2)+\beta(Q^2,\theta)G_M^{s}(Q^2)$
  as a function of $Q^2$.
The dotted fields are the $\chi$QSM-predictions for the A4
experiment at $\theta=35°$ and $\theta=145°$. The theoretical
error fields are given by assuming the Yukawa mass of the
solitonic profile in the $\chi$QSM to coincide with the pion mass
or the kaon mass, respectively. }
  \label{fig:6}
\end{figure}

\begin{figure}[h]
  \centering
\includegraphics[height=7cm]{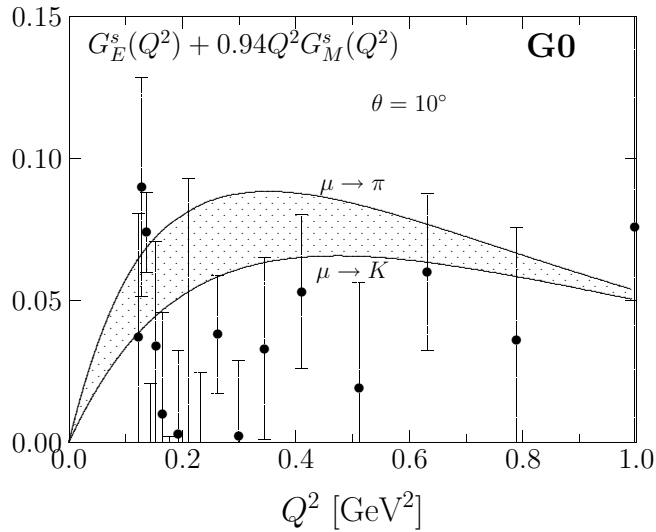}
  \caption{The values of $G_E^{s}(Q^2)+\eta G_M^{s}(Q^2)$ with
  $\eta=0.94Q^2$
  as a function of $Q^2$.
They are the predictions for the G0 experiment at $\theta=10°$.
The theoretical error field is given by assuming the Yukawa mass
of the solitonic profile in the $\chi$QSM to coincide with the
pion mass or the kaon mass, respectively. }
  \label{fig:7}
\end{figure}

\begin{figure}[h]
  \centering
\includegraphics[height=7cm]{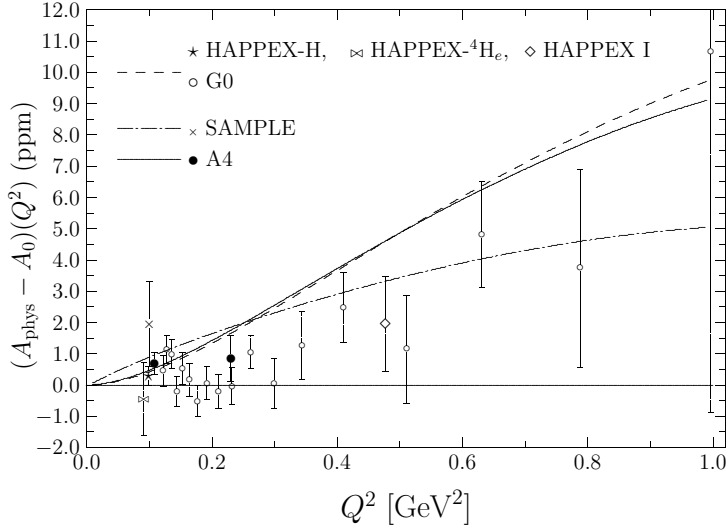}
  \caption{Difference between the parity-violating asymmetries
including strange quark effects ($A_{\rm phys}$) and the asymmetry
including just $u$ and $d$ quark contributions ($A_0$). The lines
represent the $\chi$QSM results for the kinematics (laboratory angles)
of the experiments enumerated. The curves for the small angle
forward case (G0, HAPPEX: $\theta\sim 8^\circ$) almost overlap
each other and differ slightly from A4, $\theta= 35^\circ$ (solid
line). SAMPLE is a backward angle experiment,  $\theta=
146^\circ$.} 
  \label{fig:10}
\end{figure}

\begin{figure}[h]
  \centering
\includegraphics[height=7cm]{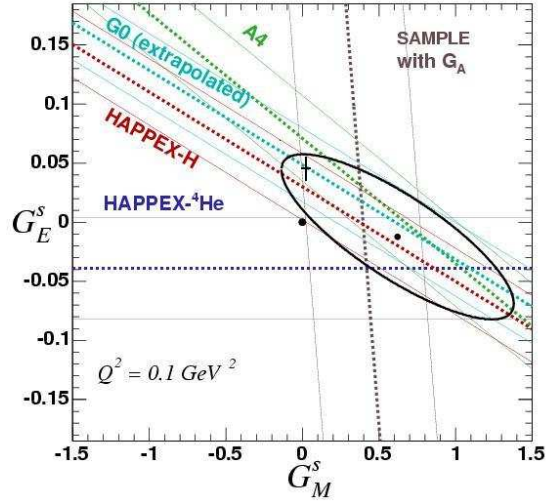}
  \caption{The world data for $G_E^s$ and $G_M^s$ from A4, HAPPEX,
SAMPLE and G0 experiments at $Q^2 = 0.1\, {\rm GeV}^2$. The plot is
taken from HAPPEX~\cite{Aniol:2005zg} and the ellipse reflects the
95 $\%$ confidence level. The theoretical number obtained by the 
$\chi$QSM is indicated by a cross which reflects the theoretical
errors. The dots indicate the center of the ellipse and the point
with vanishing strange form factors. }
  \label{fig:8}
\end{figure}

\begin{figure}[h]
  \centering
\includegraphics[height=7cm]{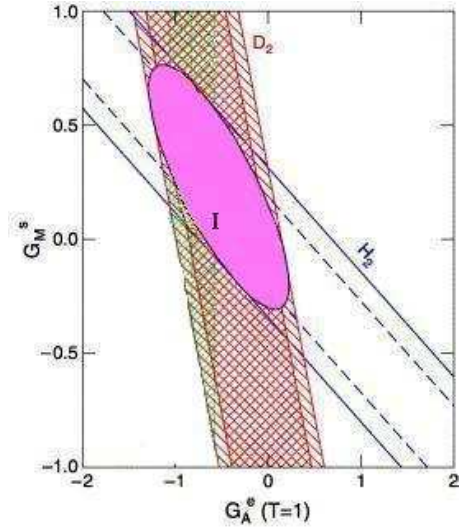}
  \caption{The hydrogen and deuterium data for $G_M^s$ and $G_A^e
  (T=1)$   from HAPPEX at $Q^2 = 0.1\,{\rm GeV}^2$. The ellipse represents the
 1 $\sigma$ overlap of the two measurements. The theoretical number
obtained by the $\chi$QSM is indicated with the bar which reflects the 
theoretical error. The data-plot is taken from Ref.
~\cite{Beise:2004py}}
  \label{fig:9}
\end{figure}

We present our numerical results in Figs.~\ref{fig:1}-\ref{fig:5}
at relevant kinematics to the A4, G0, HAPPEX, and SAMPLE
experiments in comparison with the data. The dotted curves depict
the PVA without the strange quark contribution. This means we put
$\mathcal{A}_s =0$ in Eq.~\ref{eq:asymmetry}. The dashed ones are
obtained by using the form factors from the SU(3)-$\chi$QSM
without the electroweak radiative corrections, i.e. with $\rho'$
and $\kappa'$ set equal to zero, while the solid ones ($\chi$QSM)
are our final theoretical asymmetries including those corrections.
One notices that the effect of the electroweak radiative
corrections is rather tiny. One also notices that with increasing
$Q^2$ the PVA without strange contribution deviates more and more
from the experiments, which means that with increasing $Q^2$ the
contribution of the strange quarks gets larger and larger reaching
in the end an amount up to $40\,\%$ in the present model.

As shown in Figs.~\ref{fig:1}-\ref{fig:4}, the present results are
in a good agreement with the experimental data from A4, HAPPEX,
and SAMPLE at small and intermediate $Q^2$. However, since the G0
experiments have measured the PVA over the range of momentum
transfers $0.12 \le Q^2 \le 1.0\,{\rm GeV}^2$, it is more
interesting to compare our results with them. In fact, the
predicted PVA in the present work describes remarkably well the G0
data over the full range of $Q^2$-values. It indicates that the
present model produces the correct $Q^2$-dependence of all the
form factors relevant for the PVA.

Figure~\ref{fig:5} depicts the prediction for the backward G0
experiment at $\theta = 108^\circ$ whose data are announced to be
available in near future.

\vspace{0.5cm} {\bf 4.} The Figures~\ref{fig:6}-\ref{fig:10} yield
further data which allow a detailed comparison between experiment
and theory. Fig.~\ref{fig:6} shows the typical combination
$G_E^{s}(Q^2)+\beta(Q^2,\theta)G_M^{s}(Q^2)$ playing a key role in
the experiments. In forward direction A4 has measured two points
of this observable at small $Q^2$-values, which are both well
reproduced by the $\chi$QSM calculations. The dotted error band
indicates a systematic error of the $\chi$QSM, since the soliton
is bound to have the same profile function in the up-, down- and
strange direction, see ref.~\cite{Silva:2001st} for details.
Fig.~\ref{fig:7} shows a similar combination for G0, where the
$\beta$ is assumed to be equal to $\eta =0.94\,Q^2$. In this plot
the experimental data are again resonably well reproduced by the
$\chi$QSM.

Actually one can see at Fig.~\ref{fig:8} how the $\chi$QSM values
for $G_E^s$ and $G_M^s$ fit into the present world data at
$Q^2=0.1\, {\rm GeV}^2$.  The plot is taken from
HAPPEX.~\cite{Aniol:2005zg} and the ellipse reflects the 95 $\%$
confidence level. Apparently there is good agreement beween
the $\chi$QSM and the data. A similar conclusion can be drawn from
Fig.~\ref{fig:9}, in which for $G_M^s$ and $G_E^s (T=1)$ the $\chi$QSM
is confronted with the data. Here the ellipse represents the
1-$\sigma$ overlap of the deuterium and hydrogen measurements. This
figure is taken from Beise et al.~\cite{Beise:2004py} of the HAPPEX
collaboration. 

In Fig.~\ref{fig:10} the PVAs of the various experiments
are presented focussing on the strange contribution. Following
Eq.(\ref{eq:asym}) plotted are $\mathcal{A}_{phys}
-\mathcal{A}_{0} = \mathcal{A}_{s} $. The curves are from the
$\chi$QSM. Actually the calculations  yield for the HAPPEX-experiments
and the G0-experiment nearly identical curves which cannot be
distinguished in Fig.~\ref{fig:10}.  One notes for this sensitive
quantity, originating solely from the  strange quarks of the Dirac
sea, a good agreement between theory  and experiment.

\vspace{0.5cm} {\bf 5.}  In the present work, we have investigated
the parity-violating asymmetries in the elastic scattering of
polarized electrons off protons within the framework of the chiral
quark-soliton model ($\chi$QSM).  We used as an input the
electromagnetic and strange vector form factors calculated in the
former works~\cite{Silva:2001st,Silva:2002ej,Silva}, yielding both
positive magnetic and electric strange form factors, and the
axial-vector form factors~\cite{Silvaetal} from a recent
publication. All these form factors, incorporated in the present
work, were obtained with one fixed set of four model parameters,
which has been adjusted several years ago to basic mesonic and
baryonic observables. In fact, the parity-violating asymmetries
obtained in the present work are in a remarkable agreement with
the experimental data, which implies that the present model
($\chi$QSM) produces reasonable form factors of many different
quantum numbers. We also predicted in the present work the
parity-violating asymmetries for the future G0 experiment at
backward angles. Altogether, comparing the results of the
$\chi$QSM with the overall observables of SAMPLE, HAPPEX, A4 and
G0 one observes a remarkable agreement.
\vspace{0.5cm}

The authors are grateful to Frank Maas for useful comments and
discussions.  AS acknowledges partial financial support from
Portugese Praxis XXI/BD/15681/98. The work has also been supported
by Korean-German grant of the Deutsche Forschungsgemeinschaft and
KOSEF (F01-2004-000-00102-0).  The work is partially supported by
the Transregio-Sonderforschungsbereich Bonn-Bochum-Giessen as well
as by the Verbundforschung of the Federal Ministry for Education
and Research.  The work of HCK is also supported by Korea
Research Foundation (Grant No. KRF-2003-070-C00015).



\begin{thebibliography}{99}

\bibitem{Musolf:1993tb}
  M.~J.~Musolf, T.~W.~Donnelly, J.~Dubach, S.~J.~..~Pollock,
  S.~Kowalski, and E.~J.~Beise, 
  Phys.\ Rept.\  {\bf 239}, 1 (1994).

\bibitem{Kumar:2000eq}
  K.~S.~Kumar and P.~A.~Souder,
  Prog.\ Part.\ Nucl.\ Phys.\  {\bf 45}, S333 (2000).

\bibitem{Beck:2001dz}
D.~H.~Beck and B.~R.~Holstein,
Int.\ J.\ Mod.\ Phys.\ E {\bf 10}, 1 (2001).

\bibitem{Beck:2001yx}
D.~H.~Beck and R.~D.~McKeown,
Ann.\ Rev.\ Nucl.\ Part.\ Sci.\  {\bf 51}, 189 (2001).

\bibitem{Ramsey-Musolf:2005rz}
  M.~J.~Ramsey-Musolf,
  arXiv:nucl-th/0501023.
\bibitem{Cahn:1977uu}
  R.~N.~Cahn and F.~J.~Gilman,
  Phys.\ Rev.\ D {\bf 17}, 1313 (1978).

\bibitem{Kaplan:1988ku}
D.~B.~Kaplan and A.~Manohar,
Nucl.\ Phys.\  B {\bf 310}, 527 (1988).

\bibitem{Mueller:1997mt}
B.~Mueller {\it et al.}  [SAMPLE Collaboration],
Phys.\ Rev.\ Lett.\  {\bf 78}, 3824 (1997).

\bibitem{Spayde:2000qg}
D.~T.~Spayde {\it et al.}  [SAMPLE Collaboration],
Phys.\ Rev.\ Lett.\  {\bf 84},  1106 (2000); Phys.\ Lett.\ B {\bf
583}, 79 (2004).

\bibitem{SAMPLE00s} R.~Hasty {\it et al.}  [SAMPLE Collaboration],\
Science \ {\bf 290}, 2117 (2000).

\bibitem{Aniol:2000at}
K.~A.~Aniol {\it et al.}  [HAPPEX Collaboration],
Phys.\ Lett.\  B {\bf 509}, 211 (2001).

\bibitem{Maas}
  F.~E.~Maas {\it et al.}  [A4 Collaboration],
  Eur.\ Phys.\ J.\ A {\bf 17}, 339 (2003); Phys.\ Rev.\ Lett.\  {\bf
  93}, 022002 (2004); Phys.\ Rev.\ Lett.\  {\bf 94}, 152001 (2005). 

\bibitem{happex}
  K.~A.~Aniol {\it et al.}  [HAPPEX Collaboration],
  Phys.\ Rev.\ C {\bf 69}, 065501 (2004).

\bibitem{Aniol:2005zf}
  K.~A.~Aniol {\it et al.}  [HAPPEX Collaboration],
  Phys. Rev. Lett. 96, 022003 (2006).  

\bibitem{Aniol:2005zg}
  K.~A.~Aniol {\it et al.}  [HAPPEX Collaboration],
  arXiv:nucl-ex/0506011.

\bibitem{Armstrong:2005hs}
  D.~S.~Armstrong {\it et al.}  [G0 Collaboration],
  arXiv:nucl-ex/0506021.

\bibitem{Christov:1996vm}
C.~V.~Christov {\it et al.},
Prog.\ Part.\ Nucl.\ Phys.\  {\bf 37}, 91 (1996).

\bibitem{Alkofer:1994ph}
R.~Alkofer, H.~Reinhardt, and H.~Weigel,
Phys.\ Rept.\  {\bf 265}, 139 (1996).

\bibitem{Diakonov:1996sr}
D.~Diakonov, V.~Petrov, P.~Pobylitsa, M.~V.~Polyakov, and C.~Weiss,
Nucl.\ Phys.\  B {\bf 480}, 341 (1996).

\bibitem{Petrov:1998kf}
V.~Y.~Petrov, P.~V.~Pobylitsa, M.~V.~Polyakov, I.~Bornig,
K.~Goeke, and C.~Weiss,
Phys.\ Rev.\  D {\bf 57}, 4325 (1998).

\bibitem{Goeke:2001tz}
K.~Goeke, M.~V.~Polyakov, and M.~Vanderhaeghen,
Prog.\ Part.\ Nucl.\ Phys.\  {\bf 47}, 401 (2001).

\bibitem {Diakonov:1997mm}D.~Diakonov, V.~Petrov, and M.~V.~Polyakov,
Z.\ Phys.\ A \textbf{359}, 305 (1997)

\bibitem{Silva:2001st}
  A.~Silva, H.-Ch.~Kim, and K.~Goeke,
  Phys.\ Rev.\ D {\bf 65}, 014016 (2002).
  [Erratum-ibid.\ D {\bf 66}, 039902 (2002)].

\bibitem{Silva:2002ej}
  A.~Silva, H.-Ch.~Kim, and K.~Goeke,
  Eur.\ Phys.\ J.\ A {\bf 22}, 481  (2004).

\bibitem{Silva} A. Silva, Ph.D. Dissertation (Ruhr-Universit\"at
Bochum, unpublished) (2004). 

\bibitem{Silvaetal} A.~Silva, H.-Ch.~Kim, D. Urbano, and K.~Goeke,
Phys.\ Rev.\ D {\bf 72}, 094011 (2005).

\bibitem{Eidelman:2004wy}
  S.~Eidelman {\it et al.}  [Particle Data Group],
  Phys.\ Lett.\ B {\bf 592}, 1 (2004).

\bibitem{Zhu:2000gn}
S.~L.~Zhu, S.~J.~Puglia, B.~R.~Holstein, and M.~J.~Ramsey-Musolf,
Phys.\ Rev.\ D {\bf 62}, 033008 (2000).

\bibitem{Praszalowicz:1998jm}
  M.~Praszalowicz, T.~Watabe, and K.~Goeke,
  Nucl.\ Phys.\ A {\bf 647}, 49 (1999).

\bibitem{Alberico:2001sd}
W.~M.~Alberico, S.~M.~Bilenky, and C.~Maieron,
Phys.\ Rept.\  {\bf 358} 227 (2002).

\bibitem{Beise:2004py}
  E.~J.~Beise, M.~L.~Pitt, and D.~T.~Spayde,
  Prog.\ Part.\ Nucl.\ Phys.\  {\bf 54}, 289 (2005).



\end{thebibliography}
\end{document}